# Bimodal molecular mass distribution in surfactant-free emulsion polymerization as a consequence of "coagulative nucleation"

Marta Dobrowolska and Ger J. M. Koper[*]
Department of Chemical Engineering, Delft University of Technology,
van der Maasweg 9, 2629HZ Delft, The Netherlands

**Abstract**

It is demonstrated that the often observed broadness of the molecular weight distribution obtained from latex particles synthesized by means of surfactant-free emulsion polymerization results from the multistage kinetics. The initial stage of the polymerization, by which the primary particles are formed, is of the 01-kind which means that it can be assumed that the particles are so small that at any moment of time there is no more than one radical chain per particle. After the aggregation of primary particles into secondary particles, the so called "coagulative nucleation" step, the polymerization kinetics in the subsequently coalesced secondary particles is of the pseudo-bulk kind which means that there are so many radicals in the particles that the polymerization process proceeds as in bulk. The important consequence is that the molecular weight of the 01-process is about one order of magnitude larger than that of the pseudo-bulk process. Hence, each of these polymerization modes and the aggregation stage in-between leave their traces in the molecular weight distribution as is shown by experiments aimed at prolonging the initial formation kinetics. Bimodal molecular weight distributions are found both for ionic and non-ionic initiators.

[*] Corresponding author. Tel: +31 (15) 278 8218. E-mail: G.J.M.Koper@tudelft.nl; (Ger Koper).

**Introduction**

In addition to the average molecular weight itself, an important parameter determining the applicability of polymer in latexes for various purposes is the molecular weight distribution of the polymer [1]. For instance, the width of the molecular weight distribution strongly affects the glass temperature and hence is important for all kinds of variables such as the rate of film formation or the thickness of a coating formed by the polymer material [1].

Gel permeation chromatography (GPC) is a type of size exclusion chromatography (SEC): it separates the chemical components based on their size or hydrodynamic volume. Separation occurs via the use of porous beads packed in a column. Smaller components can enter the pores more easily therefore spend more time in a column and have longer elution times [2]. The technique is akin to high-pressure liquid chromatography (HPLC), where separation is supposed to be independent of hydrodynamic volume and based on the interaction between the components and the absorbent present in the column. In practice, the distinction is not so sharp and in the case of SEC there are interaction effects of components and column material whereas in HPLC size effects are also relevant. The important realization is that in the case of SEC the obtained elution volume versus time signal varies monotonically with the molecular weight of the eluent. Hence, extensive calibration methods are used to obtain reliable information [3]. Another drawback, in particular for polymer dispersions, is that the polymer material first needs to be dissolved in a suitable solvent before a molecular weight analysis can be done [3]. This renders the technique rather cumbersome and hampers use in on-line applications where the temporal molecular weight distribution is sought for.

Quantitative prediction of – unimodal – molecular weight distributions is nowadays well possible, see for instance the work by Gilbert et al. [4,5], but has not become very popular so far. An extensive experimental investigation on the evaluation of the molecular weight distribution for styrene emulsion polymerization [5] did not invite many researchers to follow. This will at least partly be due to the fact that such predictions require a significant number of parameters that may only be known for model systems such as styrene. Also, for many applications detailed information on the formation process is not called for and a good characterization of the final product suffices. An important aspect of the model is that it predicts the tail of the molecular weight distribution to become linear in a semi-logarithmic plot of chain number versus molecular weight. The slope is characteristic of the dominant mode of chain termination [5]. In practice, however, this slope is only partially linear which could be interpreted as multimodal behaviour. Unfortunately, in such cases even more parametric information is required to describe the process [5].

There are various experimental conditions where the obtained molecular weight distribution is not unimodal and eludes quantitative predictions as discussed above. A well-known case, that does not involve emulsion polymerization, is that of ethylene where the bimodality is claimed to be due to a limited free volume effect where long chains are formed with relatively small side branches [6]. But even for the model system of styrene bimodality has been found, for instance by Tauer et al. [7]. There, multimodality is most likely due to the special conditions under which the experiments were run so that one was able to distinguish the very low molecular weight species that still resided in the aqueous phase and the higher molecular weight species from the latex particles.

Recently, we have conducted styrene emulsion polymerization using KPS and AIBN as initiator in such a way that primary polymer particles developed as a result of an almost burst nucleation [] event followed by partially reversible aggregation and subsequent coagulation into a dispersion of relatively uniform secondary particles [9-10]. Under starved monomer conditions the secondary particles were allowed to grow until polymerization was stopped by the addition of sodium nitrite ($NaNO_2$). Here we study the molecular weight distributions obtained for such emulsion polymerization experiments, which were found to be relatively broad and under some conditions exhibit bimodality.

**Materials and Methods**

*Chemicals.* Styrene (Sigma Aldrich, ≥99%) was purified by use of a pre-packed column (Sigma Aldrich) for tert-butylcatechol removal. Potassium persulfate (KPS, Sigma Aldrich) and sodium nitrite (NaNO2, Acros Organics, 98.5% pure for analysis) was used as received. Water was purified by Millipore AFS 3 water purification system.

*General procedure for polymerization.* Reactions were performed in a three-neck glass flask filled with deionized water (90 mL). To remove the oxygen and traces of other gases, water was bubbled with nitrogen and subsequently degassed under vacuum, which was repeated 3 times. Purified styrene (1.22 mL) was injected into the reaction flask and equilibrated for 2 h at 60 $^o$C. Polymerization was started by injection of 10 ml of a 50mM aqueous potassium persulfate (KPS) solution. After 180 minutes, the reaction was stopped by the addition of sodium nitrite ($NaNO_2$).

*Particle size measurements* of water-suspended samples were analysed by dynamic light scattering (DLS, Zetasizer Nano ZS, Malvern). The instrument is using 173$^o$ angle non-invasive back-scatter mode and M3-phase analysis light scattering mode, using a red 4.0 mW, 633 nm He-Ne laser. The multiple peak high-resolution fitting procedure was used to obtain the particle size distribution from the auto-correlation function.

*Gel Permeation Chromatography (GPC)* analyses were performed on a Waters system equipped with Empower software, a 515 HPLC pump, a 2410 Refractive Index detector and two Plgel 5 microns (500 A, 300 mm x 7.5 mm) columns with THF as solvent at a flow rate of 1 mL/min at room temperature. All samples were prepared with a concentration of 0.05g polymer/10 mL THF. Results are plotted on a semi-logarithmic graph of the weight averaged molecular weight density d(wt)/d(log Mw) versus nominal molecular weight of the almost monodisperse calibration samples.

*Cryo Transmission Electron Microscopy.* Pictures were obtained on a Philips CM120 electron microscope operating at 120 KV. Samples were prepared by depositing a few microliters of suspension on a holey carbon coated grid (Quantifoil 3.5/1, Quantifoil micro tools GmbH, Jena, Germany), after blotting away the excess of liquid the grids were plunged quickly in liquid ethane. Frozen hydrated specimens were mounted in a cryo specimen holder (Gatan, model 626). Micrographs were recorded under low dose conditions on a slow scan CCD camera (Gatan, model 794).

**Results and Discussion**

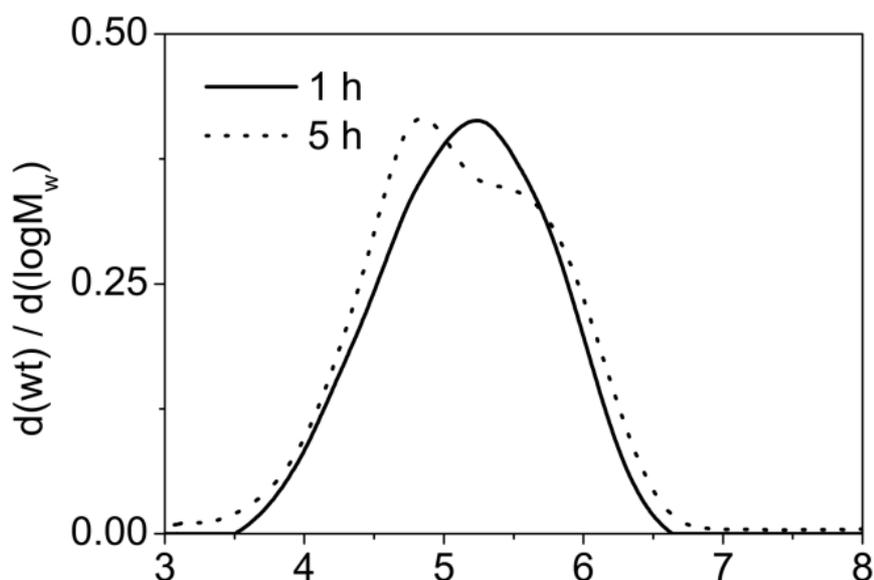

**Figure 1.** Early (1h) and late (5h) stage molecular weight distribution for polystyrene initiated with 2.5mM KPS.

As discussed above, the experiments were conducted in such a way as to optimize the visibility of the aggregation of primary particles into secondary particles. The primary particles were formed by burst nucleation after the introduction of the initiator in the initial stage. In this stage, the concentration of monomer in the aqueous phase is, due to the stirring, significantly higher than the equilibrium styrene solubility at that temperature which is signalled by a low concentration of droplets visually observed by a low turbidity of the mixture. Diffusion of monomer from the styrene top phase is very slow compared to the consumption by the polymerization process so that the rest of the polymerization is performed under starved monomer conditions. This allows one to study the reversible aggregation and subsequent coagulation into secondary particles in detail.

During and immediately after nucleation, the particles are still very small and chain growth will largely be of the 01-kind which means that due to the confinement within which polymerization takes place there are only very few active radical chains of which the kinetics can be approximated by a description that assumes the presence of none or just one radical chain, see e.g. [4] and references therein. In contrast, after coagulation of the primary particles into the significantly larger secondary particles there will be so many free radical chains that the growth is of the pseudo-bulk (pb) kind, i.e. that chain growth within the particles can be approximated by a description that assumes the reaction kinetics to be similar to the situation in bulk. Distinctive phenomena of the two extreme kinds of polymerization kinetics are, that under otherwise similar conditions in the 01-case the reaction rate is almost an order of magnitude faster [5]. In particular, chain termination is very different in 01-kinetics compared to pb-kinetics and as a consequence, the average molecular weight achieved by 01-kinetics is about one order of magnitude larger than what is achieved by pb-kinetics. Important to note is that this difference in average molecular weight is already attained at low levels of conversion above which only the value for the 01-case still increases whereas the average remains constant for the pb-case [5]. The resulting molecular weight distributions as depicted in Figure 1 clearly illustrate the above described phenomenon where it is important to note that [9] it was illustrated by means of high resolution scanning electron micrographs, see Figure 2.

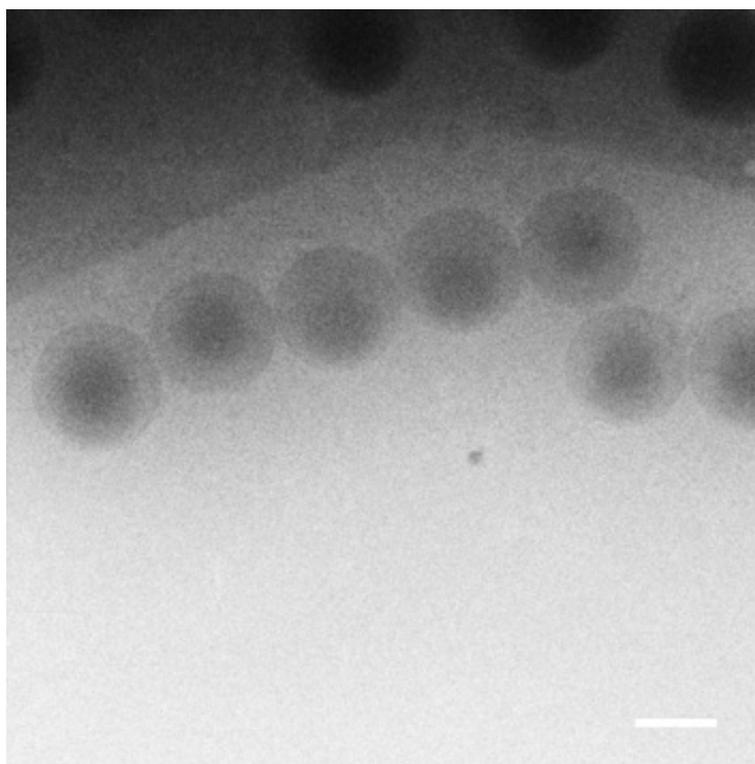

**Figure 2.** Cryo-TEM image of polystyrene particles after 5h polymerization time (scale bar is 200nm).

At early stages, the distinct primary particles are clearly noticeable whereas at late stages their structure is almost completely smeared out. Hence, Figure 1 can be taken to illustrate that at early stages of polymerization the kinetics is of the 01-kind and takes place largely within the primary particles, whereas at late stages the kinetics is as in bulk. As the coagulation of primary particles into secondary particles involves an enhancement of termination events of active chains from neighbouring primary particles, termed *coagulative termination* by Herrera-Ordonez et al. [11], an additional contribution at about a factor $\sqrt{2}$ larger average molecular weight achieved during the early polymerization stage is expected. This explains the broadening that is observed at later stages for the higher molecular weights.

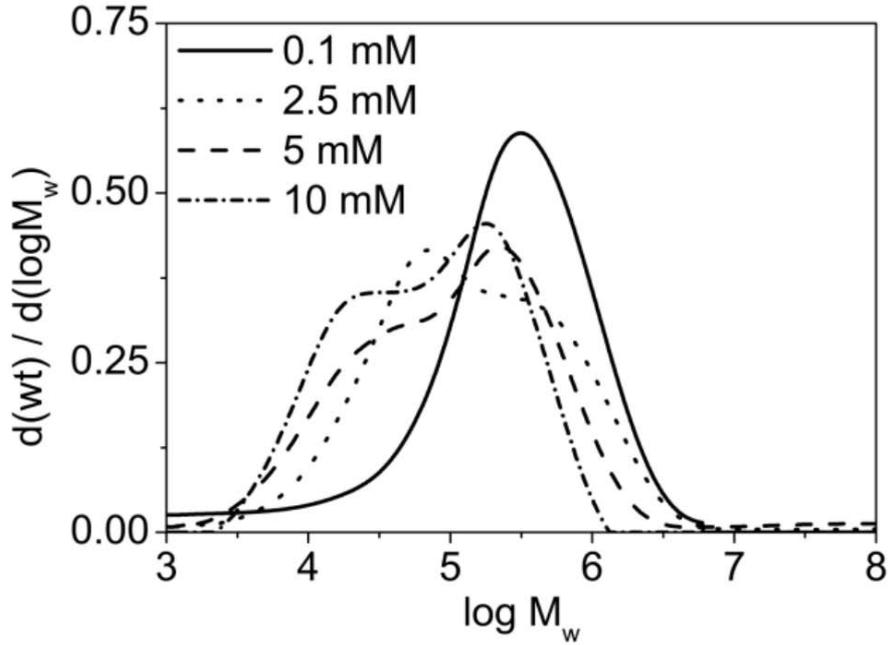

**Figure 3.** Molecular weight distribution for polystyrene initiated with 0.1mM; 2.5mM; 5mM and 10mM KPS (all the measurements were taken after 5h polymerization time).

The dependence on initiator concentration, as depicted in Figure 3, is the more interesting as more processes with different time scales leave their traces in the molecular weight distribution. As described above, the development during the early stages of polymerization is reflected by the high molecular weight part of the distribution. We concentrate here in particular on the right flank of the distributions of Figure 3. The initiator concentration dependence of this part of the distribution clearly indicates that the average molecular weight for the primary particles decreases with increasing initiator concentration which is consistent with the general behaviour of the 01-model for polymerization kinetics, see e.g. [4]. From the low molecular weight regime of the distributions, the behaviour at later polymerization stages can be deduced. We concentrate here in particular on the left flank of the distributions of Figure 3. It is observed that the higher the initiator concentration, the lower the average molecular weight for the pseudo-bulk polymerization at the later stages in the larger secondary particles. For the lowest initiator concentration of the two contributions cannot be distinguished. The intricate interplay between these processes, polymerization in the primary particles, the coagulation process, and subsequent growth in the secondary particles makes the middle part of the molecular weight distributions rather confusing with respect to their initiator concentration dependence. This is why we concentrated on the left and right parts of the distribution, from which the behaviour is sufficiently clear.

In Figure 4 the particle size, as measured by photon correlation spectroscopy, is plotted versus time. For late stages, the dependence is practically linear in time with a slope that is formally given by [12]

$$\frac{da}{dt} = \frac{M}{\rho} D \frac{c_s}{\delta} \qquad (1)$$

The equation follows from the combination of mass conservation and Fick's diffusion equation where the concentration gradient is estimated as the jump value across the hydrodynamic boundary layer. As this limiting behaviour is expected for the starved polymerization regime, the diffusing species is considered to be single styrene molecules with molar mass M = 104 g/mol and a diffusion coefficient in water of about D = 8 ·10$^{-6}$ cm$^2$/s. The mass density of the particles is about that of polystyrene, i.e. ρ = 0.96-1.04 g/cm³, and the solubility of styrene in water is $c_s$ ≈ 5 mol/m³. Estimating the hydrodynamic boundary layer thickness for the slow stirrer speed of 150 rpm at δ ≈ 0.6 mm using the Blasius solution, we obtain a value of about 0.7 nm/min as observed in Figure 4.

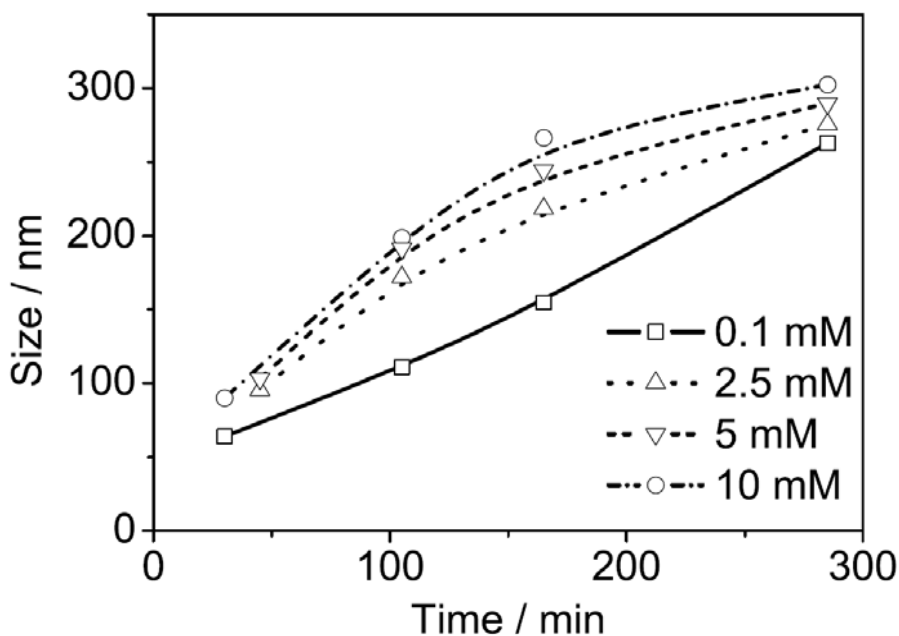

**Figure 4.** Particle size as a function of polymerization time for different initiator concentrations. Lines indicate regressions of the late stage behaviour.

The lowest initiator concentration results are, within the investigated time scale of 5 h, apparently not yet exhibiting the limiting behaviour. The slight variations in slope that are found for the higher initiator strengths most probably reflect the slight dependence of the hydrodynamic boundary thickness on particle size.

The growth of the primary particles will not be visible from Figure 4, but the coagulation into secondary particles and the subsequent growth is what is visible. The initial slopes increase with increasing initiator concentration because more initiator will mean more primary particles to be formed. A higher concentration of primary particles implies a faster growth of the secondary particles. However, the final particle size hardly depends on initiator concentration but only depends on the growth under starved conditions by means of polymerization of the secondary particles.

An interesting question that remains to be answered is whether the observed behaviour is special to the ionic initiator KPS or whether under otherwise similar circumstances also other initiators leave the history of their evolution so clearly in the molecular weight distribution. In a previous manuscript we have demonstrated that with the non-ionic initiator AIBN quite similar results could be obtained as with KPS provided that the pH was in the neutral regime and that the ionic strength was near the optimal value of about 1 mM. In Figure 5 it is illustrated that the answer to the above question is probably positive and that indeed also here but less clear a bimodal molar weight distribution can be discerned.

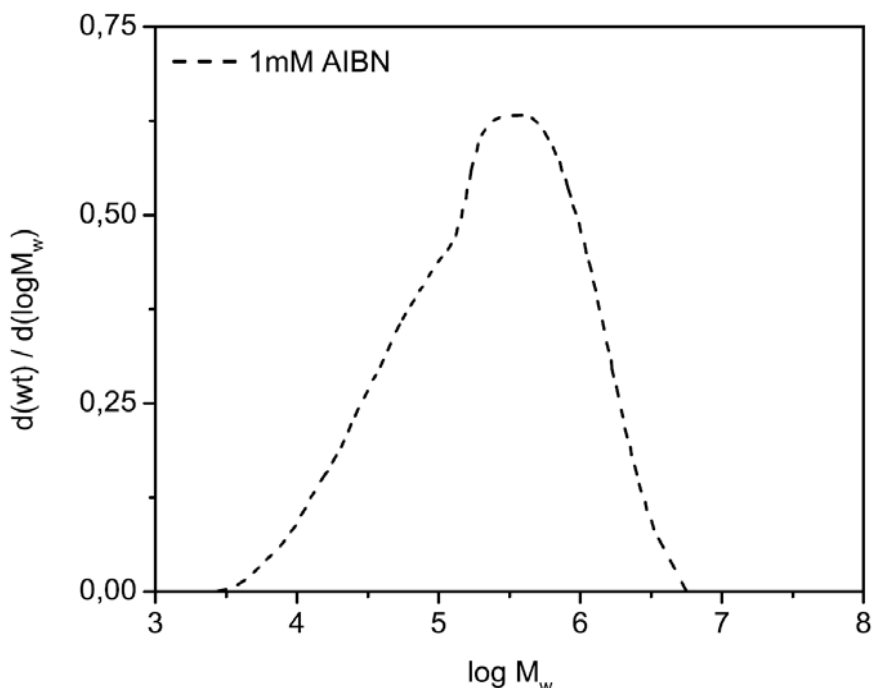

**Figure 5.** Molecular weight distribution for polystyrene initiated with 1 mM AIBN, measurement was taken after 5 h polymerization time.

**Conclusion**

The above-described experiments were designed to demonstrate particular aspects about the early stage kinetics of surfactant free emulsion polymerization. In previous work we have demonstrated that using this scheme the coagulative nucleation process can be clearly visualized for the case of ionic initiation, and in a subsequent contribution we demonstrated this for a completely different, non-ionic, initiator. Here, we demonstrate that the initial stage of coagulative nucleation, the primary particles, and the final stage, the secondary particles, have a different mode of polymerization kinetics that can be characterized as 01-kinetics and pseudo-bulk kinetics respectively. We demonstrated that these two different modes leave their clear signature in the molecular weight distribution and render these bimodal. We surmise that the archetypical characterization of broad molecular weight distributions to emulsion polymerization processes in actual fact is due to the fact that in many cases these different modes contribute with various intensities and separation in molecular weight. Hence, the signature is actually a characterization of the mix of acting kinetic mode.

**Acknowledgement**

Martin Fijten (Technische Universiteit Eindhoven) is acknowledged for the GPC analysis.

# Appendix

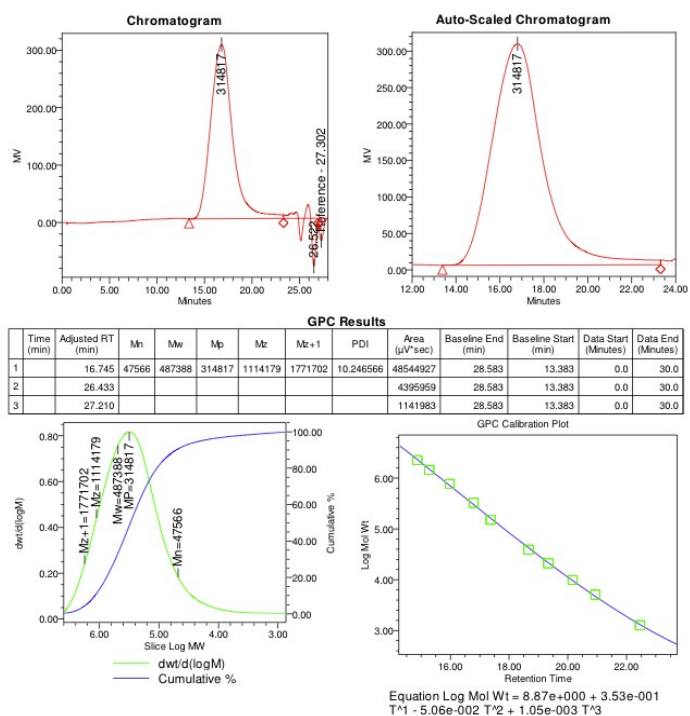

**Figure S1.** GPC method report for polystyrene with 0.1mM KPS as initiator.

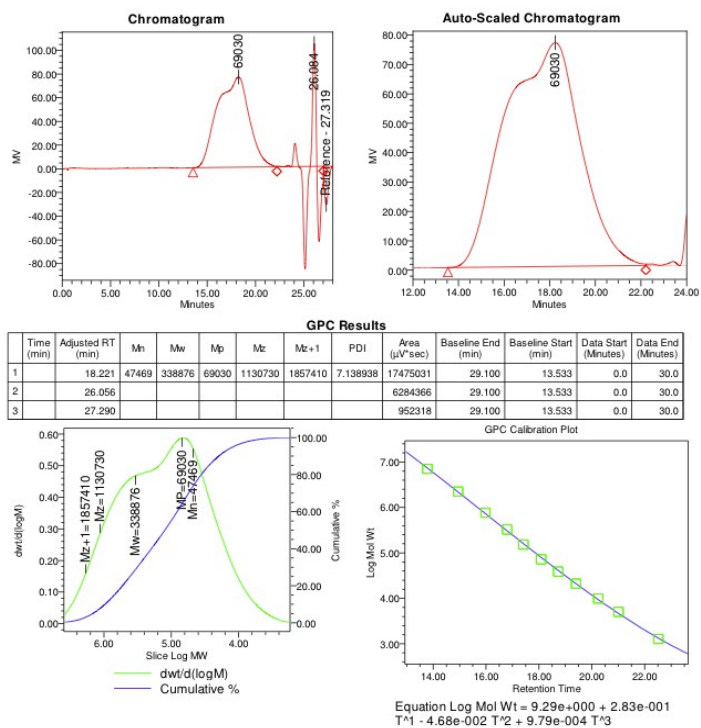

**Figure S2.** GPC method report polystyrene with 2.5mM KPS as initiator.

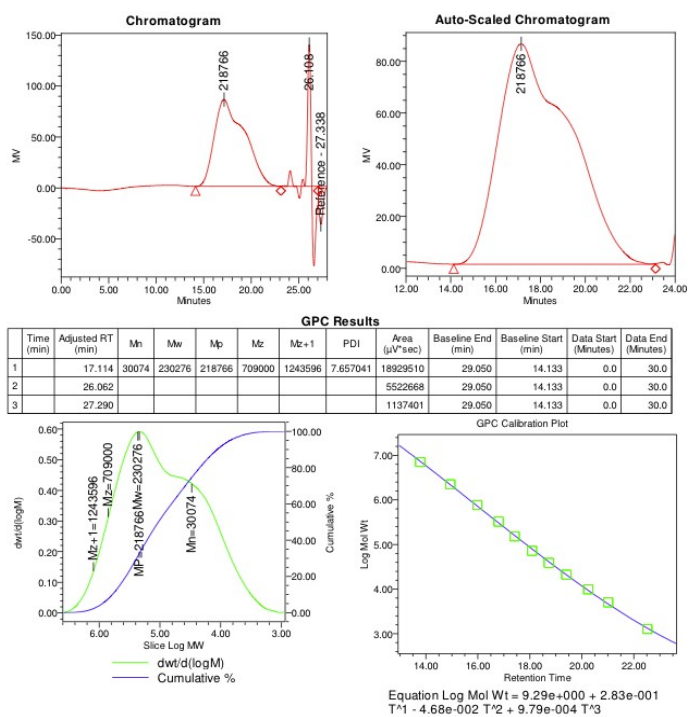

**Figure S3.** GPC method report polystyrene with 5mM KPS as initiator.

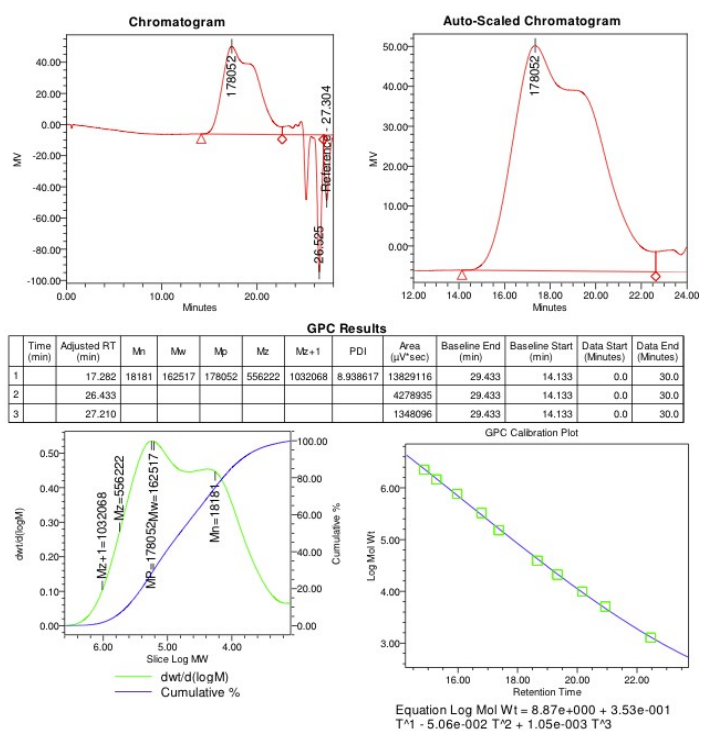

**Figure S4.** GPC method report polystyrene with 10mM KPS as initiator.

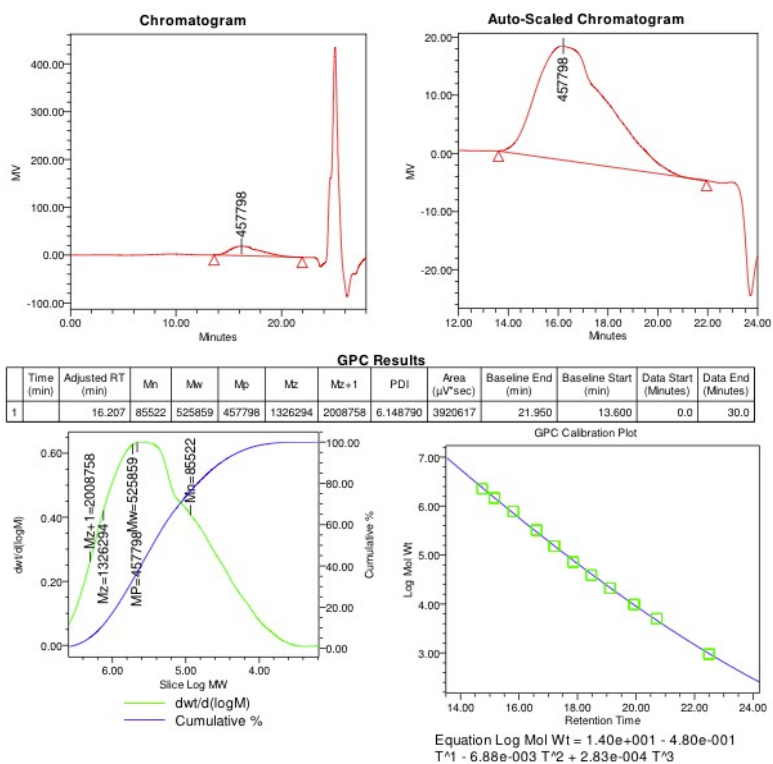

**Figure S5.** GPC method report polystyrene with 1mM AIBN as initiator.